\documentclass[11pt]{amsart}
\usepackage{graphics}
\usepackage{hyperref}
 \font\p=cmr9 
\def \dd{{\rm d}}
\def \DD{{\rm D}}
\numberwithin{equation}{section}
\begin{document}
\title[Nontrivial Killing groups in general relativity] {
Interpreting solutions with nontrivial Killing groups in
general relativity}%
\author{Salvatore Antoci}%
\address{Dipartimento di Fisica ``A. Volta'' and C.N.R., Pavia, Italia}%
\email{salvatore.antoci@unipv.it}%
\author{Dierck-Ekkehard  Liebscher}%
\address{Astrophysikalisches Institut Potsdam, Potsdam, Deutschland}%
\email{deliebscher@aip.de}%

\begin{abstract}
General relativity is reconsidered by starting from the
unquestionable interpretation of special relativity, which (Klein
1910) is the theory of the invariants of the metric under the
Poincar\'e group of collineations. This invariance property is
physical and different from coordinate properties. Coordinates are
physically empty (Kretschmann 1917) if not specified by physics,
and one shall look for physics again through the invariance group
of the metric. To find the invariance group for the metric, the
Lie ``Mitschleppen'' is ideal for this task both in special and in
general relativity. For a general solution of the latter the
invariance group  is nil, and general relativity behaves as an
absolute theory, but when curvature vanishes the invariance group
is the group of infinitesimal Poincar\'e ``Mitschleppen'' of
special relativity. Solutions of general relativity exist with
invariance groups intermediate between the previously mentioned
extremes. The Killing group properties of the static solutions of
general relativity were investigated by Ehlers and Kundt (1964).
The particular case of Schwarzschild's solution is examined, and
the original choice of the manifold done by Schwarzschild in 1916
is shown to derive invariantly from the uniqueness of the
timelike, hypersurface-orthogonal Killing vector of that solution.
\end{abstract}
\maketitle {}

\section{Introduction}\label{a}
The year 2010 lies, in physics, midway between two centenaries. At
the time of writing, five years have elapsed since the centenary
of the "Einsteinjahr", with the onset of special relativity theory
\cite{Einstein1905}, and five years too separate us from the
centenary of the discovery of the general theory of relativity
\cite{Einstein1915c}, \cite{Hilbert1915}. Both theories are in
present times rightly studied and considered like pillars of
physical knowledge. It is quite surprising to remark, then, that
the mutual relationship between the two theories is still clouded
by some fog that needs to be completely dissipated. This happens
because, while the physical interpretation of the special theory
of relativity is unique and clear since the writing, in 1910, of
the well known sentence by Felix Klein\footnote [2]{``Was die
modernen Physiker {\it Relativit\"atstheorie} nennen, ist die
Invariantentheorie des vierdimensionalen Raum-Zeit-Gebietes, $x$,
$y$, $z$, $t$ (der Minkowskischen ``Welt'') gegen\"uber einer
bestimmten Gruppe von Kollineationen, eben der
``Lorentzgruppe''\cite{Klein1910},\cite{Klein1911}. English
translation: ``What the modern physicists call {\it theory of
relativity} is the theory of the invariants of the space-time
region, $x$, $y$, $z$, $t$ (the Minkowski ``world'') with respect
to a given group of collineations, namely the ``Lorentz group''.},
the group theoretical interpretation of general relativity theory
is still in want of further understanding of its physical meaning
along the path set by the very early remarks of Kretschmann
\cite{Kretschmann1917} and of Noether \cite{Noether1918}
respectively. This problematic situation becomes evident whenever
solutions of that theory are encountered, that happen to possess a
nontrivial Killing structure. Like we shall see in the following,
this is a frequently overlooked, but by no means minor issue. In
fact, for ease of computation, nearly all the solutions to the
field equations of general relativity studied up to now by the
relativists are endowed with some symmetry, i.e. their metric
manifolds possess nontrivial Killing groups of invariance. Whether
and how the occurrence of these nontrivial physical structures
influences the physical interpretation of particular solutions
will be the subject of the present investigation. The weight of
the argument is best understood if one takes one step back, and
starts the discussion from the nontrivial Killing structure
associated with the invariance of the Minkowski metric manifold
under the group of collineations mentioned by Klein, whose
physical content was clear in 1910 and is still unquestionable
today.\par

Although it is not immediately prominent, the discussion of the
Killing structure is an extension of the good old principle that
rest can be defined only with respect to some other object - to
some other body in mechanics, to some other object in general. In
classical mechanics, a body can be at rest with respect to itself,
then it is inertially moving, no acceleration involved. It can be
temporarily or permanently at rest with respect to some other body
with no such condition. In Special Relativity, being at rest the
one to the other requires for both inertial, acceleration-free
motion because of the lack of absolute simultaneity. In GRT, the
two notions of being at rest with repect to one-self and being at
rest to another body fall apart. To be at rest with respect to
one-self remains inertial motion. To be at rest with respect to
another body implies to be at rest with respect to the structures
of the gravitational field produced by this second body. A body is
at rest in a gravitational field when the field does not change
along its world-line. When such a world-line exists, it is a
Killing trajectory. As we shall see, due to the more complicated
structure of a metric field, the field itself must reflect the
inertial state of its source. In the most general case, when the
Killing group is trivial, the events of space-time may be
intrinsically characterized through the different invariants to be
constructed with the Riemann tensor, and any world-line passes
through varying environments so that it can never be called to be
at rest. The gravitational field must allow a congruence of
timelike world lines, along which the field does not change. A
particle moving on such a world line can then be called at rest
with respect to the field. In other words, a rest can be defined
in a manifold with a timelike Killing congruence. These fields
will be studied in section \ref{e} . Due to the existence of the
gravitational field, being at rest is no more an acceleration-free
state. To be at rest in a gravitational field implies to be
accelerated all the time, and the acceleration is the curvature of
the Killing trajectories. When the Killing congruence is uniquely
defined, the curvature of its lines yields again a measurable
scalar quantity, as we shall discuss in section \ref{e}.

\section{The Minkowski metric in curvilinear
coordinates}\label{b} The world-lines of free motion and all the
other trajectories of the subgroup of translations of the
Poincar\'e group form a set of straight lines, which is used to
introduce the ``Galilean coordinates''\cite{LL1970} we are used
to. The metric expresses the full Poincar\'e group, and obtains
the known form in Galilean coordinates. In GRT, the world lines of
free motion do not form a set of straight lines any more, and the
invariance group may be reduced to even the trivial group. Hence,
no simplifiying coordinates exist.\par

No doubt, if the Minkowski ``world'' is described by availing of
``Galilean coordinates''\cite{LL1970} $x$, $y$, $z$, $t$, with
respect to which the metric $g_{ik}$ reads
\begin{equation}\label{2.1}
g_{ik}=\eta_{ik}=\rm{diag}(-1,-1,-1, 1),
\end{equation}
a great simplification occurs. When this representation is
adopted, the coordinates $x^i$ are not just labels for identifying
events. Due to this particular form of $\eta_{ik}$, Galilean
coordinates have a direct metric reading, i.e. to each particular
system of coordinates a physically admissible, inertial reference
frame, to be built with rods, clocks and light signals, is
directly associated in one-to-one correspondence. Moreover, when
this representation is adopted, one recognizes that the Poincar\'e
group of transformations, besides being endowed with direct
physical meaning, is the group of invariance of $\eta_{ik}$. The
invariance of $\eta_{ik}$ under the Poincar\'e group constitutes
what Klein and later Kretschmann once called the physically
meaningful ``relativity postulate'' of the original theory of
relativity of 1905. Although Galilean coordinates are adopted for
ease of representation, it is clear that the ``relativity
postulate'', i.e. the group of invariance, is physical and
coordinate independent. Therefore other representations, that do
away from Galilean coordinates, could be availed upon as well.\par
 To ease the comparison of special relativity with the general
relativity of 1915, where the adoption of general curvilinear
coordinates, with the associated group of covariance, in keeping
with the fundamental work by Ricci and Levi-Civita \cite{RLC1900},
 is {\it de rigueur}, it is necessary to describe
 the Minkowski metric manifold in
general curvilinear coordinates too. In this way the duplicity
characteristic of the Galilean coordinates disappears. In the
Minkowski manifold curvilinear coordinates are just labels, devoid
of physical meaning beyond the mere topological identification of
the events. In fact the only physical restriction on curvilinear
coordinates, needed for preserving the individuality of the single
event, is just that any transformation between two such systems of
coordinates needs to be one-to-one. The elements of the abstract
Poincar\'e group have in general no global representation through
a coordinate transformation occurring between two arbitrarily
given systems of curvilinear coordinates $x^i$ and $x'^i$. Since
the existence of the elements of the Poincar\'e group, meant in
the abstract sense, does not depend on the choice of the
coordinates, it is fundamental to learn what subgroup of the
abstract Poincar\'e group, if any, can find a mathematically
affordable representation with respect to a general system of
coordinates. If this question is positively answered, studying the
symmetries of a pseudo-Riemannian manifold with a unique
mathematical formulation of general character that applies both
whether the curvature tensor $R_{iklm}$ of the metric manifold is
vanishing or not will become a well defined problem.\par It is
clear that curvilinear coordinates are unsuitable in general for
providing a global account of the symmetry properties of a curved
manifold. If the curvature tensor $R_{iklm}$ is nonvanishing one
shall restrict the study of these symmetries to an infinitesimal
neighbourhood of a given event. Happily enough, in this limit the
powerful mathematical tool of Lie's infinitesimal ``dragging
along'' (``Mitschleppen'' \cite{Schouten1954}) of a metric field
can be used \cite{LL1970}.\par Let us consider a pseudo Riemannian
manifold equipped with two curvilinear coordinate systems $x'^{i}$
and $x^{i}$ ($i=1,..,4$) such that
\begin{equation}\label{2.2}
x'^{i}=x^{i}+\xi^{i},
\end{equation}
where $\xi^i$ is an infinitesimal four-vector. Under this
infinitesimal coordinate transformation, the components of the
metric tensor $g'^{ik}$ in terms of $g^{ik}$ read
\begin{equation}\label{2.3}
g'^{ik}(x'^{p})=\frac{\partial x'^{i}}{\partial
x^{l}}\frac{\partial x'^{k}}{\partial x^{m}}g^{lm}(x^p) \approx
g^{ik}(x^{p})+g^{im}\frac{\partial \xi^{k}}{\partial x^{m}}
+g^{km}\frac{\partial \xi^{i}}{\partial x^{m}}.
\end{equation}
The quantities in the first and in the last term of (\ref{2.3})
are calculated at the same event (apart from higher order
infinitesimals). We desire instead to compare the quantities
$g'^{ik}$ and $g^{ik}$ calculated for the same coordinate value,
i.e. evaluated at neighbouring events separated by the
infinitesimal vector $\xi^i$. To this end, let us expand
$g'^{ik}(x^{p}+\xi^p)$ in Taylor's series in powers of $\xi^p$. By
neglecting higher order infinitesimal terms, we can also
substitute $g^{ik}$ for $g'^{ik}$ in the term containing $\xi^i$
of the expansion truncated at the first order term, and we find:
\begin{equation}\label{2.4}
g'^{ik}(x^{p})=g^{ik}(x^{p})+g^{im}\frac{\partial
\xi^{k}}{\partial x^{m}} +g^{km}\frac{\partial \xi^{i}}{\partial
x^{m}}-\frac{\partial g^{ik}}{\partial x^m}\xi^m.
\end{equation}
But the difference
$\delta{g^{ik}(x^p)}=g'^{ik}(x^{p})-g^{ik}(x^{p})$ has tensorial
character and can be rewritten as
\begin{equation}\label{2.5}
\delta{g^{ik}(x^p)}=\xi^{i;k}+\xi^{k;i}
\end{equation}
in terms of the contravariant derivatives of $\xi^i$. When
\begin{equation}\label{2.6}
\xi^{i;k}+\xi^{k;i}=0
\end{equation}
$\delta{g^{ik}(x^p)}=0$, and the metric tensor $g^{ik}$ goes into
itself under Lie's ``Mitschlep\-pen''\cite{Schouten1954} along
$\xi^i$. An infinitesimal Killing vector is defined as a
four-vector $\xi^i$ that fulfills (\ref{2.6}). We assume for
instance that, at a given event, $n$ infinitesimal Killing vectors
${_a}\xi^i, a=1,..,n$ exist. They define the infinitesimal
``Mitschleppen'' group of rank $n$, against which the metric
$g_{ik}$ remains invariant. In another instance, let us consider
the solutions of Eqs. (\ref{2.6}) that hold when $R_{iklm}=0$. Its
vectors $\xi^i$ define the elements of the infinitesimal
Poincar\'e group. Eqs. (\ref{2.6}) thereby provide the sought-for
unique mathematical description of the local symmetries for both
the special and the general theory of relativity through the
corresponding Killing groups.
\section{Kretschmann's objection to Einstein's interpretation of general
relativity}\label{c} Despite the warning implicit in Klein's
ironic sentence of 1910 \cite{Klein1910}, when, at the end of the
year 1915, both Einstein and Hilbert arrived at the field
equations of general relativity, both of them thought that their
fundamental achievement entailed, inter alia, the realisation of a
theory of gravitation whose underlying group was the group of
general coordinate transformations. At variance with Hilbert's
standpoint, that the adoption of general coordinates was per se a
great advance in physics, due to the extraordinary achievement
thereby obtained in the mathematical structure of the theory, in
Einstein's original idea the newly acquired group-theoretical
property of general covariance was believed to be an essential one
from a physical point of view. According to Einstein's original
conception of general relativity \cite{Einstein1916}, giving up
the Galilean coordinates and the a priori Minkowski metric
$\eta_{ik}$ and admitting general, curvilinear coordinates might
allow, on physical grounds, the introduction of reference frames
that do away from the arbitrary singling out of the inertial
frames as the only admissible ones. Since, according to the early
version of the equivalence principle, gravity and acceleration of
a test particle had to be identified locally at any given event,
in the newborn theory of gravitation curvilinear coordinates
should be introduced not just for availing of the convenient
mathematical tools introduced by Ricci and Levi-Civita
\cite{RLC1900}, but due to a cogent physical reason in the first
place. It is mathematically exhibited by the shifty role of
inertial and of gravitational forces, identified as the two
nontensorial addenda that appear in, say, the equation of geodesic
motion
\begin{equation}\label{3.1}
\frac{\DD^2 x^i}{\dd s^2}\equiv\frac{\dd^2 x^i}{\dd
s^2}+\Gamma^i_{kl}\frac{\dd x^k}{\dd s}\frac{\dd x^l}{\dd s}=0,
\end{equation}
and were interpreted \cite{Einstein1916} by Einstein just like
acceleration and gravitation respectively. In this way, the
vanishing of acceleration or the vanishing of gravitation at a
given event could be produced in principle through suitably chosen
coordinate transformations belonging to the group of general
coordinate transformations. In the same year 1915, however, Erich
Kretschmann had published in {\it Annalen der Physik} a long
article \cite{Kretschmann1915}, entitled ``\"Uber die prinzipielle
Bestimmbarkeit der berechtig\-ten Bezugssysteme beliebiger
Relativit\"atstheorien'', in which an accurate analysis of the
relation between observation and mathematical structure in a
theory possessing a generic postulate of relativity is developed.
No wonder then, if two years later, with the paper
\cite{Kretschmann1917} entitled ``\"Uber den physikalischen Sinn
der Relativit\"atspostulate; A. Einsteins neue und seine
urspr\"ungliche Relativit\"atstheorie'', the same author produced
an analysis of the relation between the ``special'' and the
``general'' theory of relativity that defied the previously quoted
group-theoretical assessment by Einstein, and proposed an
alternative of his own, whose objection was entirely in keeping
with Klein's remarks. The validity in principle of Kretschmann's
objection was soon acknowledged by Einstein himself
\cite{Einstein1918}. Acceptance was allotted henceforth to
Kretschmann's way of assessing the very meaning of ``relativity''.
In keeping with Kretschmann and Klein, the ``relativity content''
of a given theory should not be ascertained through the group of
covariance allowed by the particular expression adopted for
writing the equations of that theory in terms of certain
coordinates. It should be assessed through its group of
invariance, meant to be ``a physical property of the system''. As
learned in the long time elapsed since the publication of
Kretschmann's paper of 1917, in a Riemannian metric manifold the
group of invariance of the metric is directly inscribed by the
Killing vectors in the intrinsic, geometric structure of a
manifold.\par

\section{Interpretation of solutions with nontrivial Killing groups in general
relativity}\label{d} In keeping with Kretschmann's objection,
curvilinear coordinates used to describe a certain solution of
general relativity are physically vacuous, because the field
equations of any theory could be written with such coordinates,
while the invariant properties of the metric, accounted for by the
associated Killing group, convey information on the physical
content of the solution under question. The long known relation
between invariance and true conservation of physical quantities,
first investigated by Noether \cite{Noether1918}, is another proof
of the validity of the latter assertion.\par The seemingly obvious
objection raised by Kretschmann has far-reaching consequences.
First of all, while the Killing group of the metric of special
relativity is just the Poincar\'e group in the limit case of
infinitesimal motions, for a general solution of the field
equations of general relativity the Killing group reduces to the
identity, i.e. general relativity, despite its very name, in that
case behaves indeed like an absolute theory. Moreover in general
relativity particular solutions exist too, whose Killing group
happens to be intermediate between the Poincar\'e group for
infinitesimal motions, that prevails when $R_{iklm}=0$, and the
trivial group that holds when the solution under question has no
symmetry whatsoever. An overlooked property is thereby emphasized:
the general relativity of 1915 is more appropriately considered to
be a theory whose intrinsic ``relativity content'' is not given a
priori once and for all, like it happens in the special relativity
of 1905. Its true content can be ascertained only in a case by
case way, after its solutions are found, and the elements of the
Killing group are determined by solving the Killing equations
(\ref{2.6}). Different solutions, i.e. different manifolds can and
do exhibit a different relativity postulate, possibly a vanishing
relativity content. Moreover, different submanifolds of a given
manifold can and sometimes do exhibit Killing groups with a
different relativity postulate in which, according to the well
known results by Noether \cite{Noether1918}, different
conservation laws can and do prevail.\par
\section{The peculiar Killing group of the static solutions
of general relativity}\label{e} A physically quite relevant
example of solutions to the field equations of general relativity
with a nontrivial Killing group is given by the so-called static
solutions. The notion of staticness as an intrinsic feature
independent from coordinates was clearly in the mind of a
mathematician like Levi-Civita when he introduced and discussed at
length static solutions given in symmetry-adapted static
coordinates, shown in his ground-breaking work
\cite{Levi-Civita1917a} on ``Einsteinian statics'' and in the
series of eight Notes
\cite{Levi-Civita1917b}-\cite{Levi-Civita1919}, all entitled
``Einsteinian $\dd s^2$ in Newtonian fields''. The direct,
intrinsic definition of staticness through the perusal of the
``static Killing group'' only appeared much later in the chapter
published in 1964 by Ehlers and Kundt \cite{EK1964}. We shall
follow the intrinsic definition of staticness, more precisely, the
definition of {\it static vacuum fields} through their Killing
vectors, since it is the only way to enlighten a property of
uniqueness, that was given explicitly in \cite{EK1964} for the
first time, and is crucial for grasping the quite novel physical
property of staticness as it occurs in general relativity. It is
exhibited by the exact vacuum solutions, but its peculiarity is
usually paid scarce attention in the literature. In \cite{EK1964},
after having proved\par\medskip \noindent Theorem 2-3.1: ``A
space-time is static if and only if it admits a group $G_1$ of
isometries whose trajectories form a time-like, normal
congruence.''\par\medskip\noindent on page 65 Ehlers and Kundt
reach theorem 2-3.2 and the crucial\par\medskip\noindent Corollary
1. ``In a static space-time there exist precisely one static
congruence, and precisely one $G_1$ with time-like,
hypersurface-orthogonal trajectories provided that either the
conform tensor is non-degenerate or the time-like eigenvector of
the Ricci tensor is simple.''

From these results it transpires that the notion of staticness in
general relativity has a deep meaning that directly stems from the
peculiar structure of the Killing group decided by solutions to
the field equations, and finds no counterpart in the different
Killing structure of the infinitesimal Poincar\'e group set a
priori in special relativity.\par In the latter theory, through a
given event, an infinite number of distinct timelike Killing
vectors can be drawn. To any such vector, through a given event
one and just one spatial hypersurface can be found, that is
orthogonal to the chosen timelike Killing vector. This means that
the foliation of spacetime in space and time can be performed in
infinite ways in special relativity, i.e. it has no intrinsic
character \cite{Minkowski1909}. Through an infinitesimal
Poincar\'e transformation, or through a sequence of these
transformations, any one of the distinct timelike Killing vectors
can be brought to rest in the coordinate and in the reference
frame sense. This occurs in keeping with the very notion of
relativity of motion that prevails in special relativity: no
absolute rest can be defined in an intrinsic way in the Minkowski
metric manifold.
\par
According to Corollary 1, the opposite is true for the ample class
of vacuum solutions of general relativity that are named static
after Ehlers and Kundt. Through a given event of such a solution,
provided that either the conform tensor is non-degenerate or the
time-like eigenvector of the Ricci tensor is simple, a unique
timelike Killing vector exists, that is hypersurface-orthogonal
too. From it, one and just one static congruence is defined
through the given event. When this is the case, the foliation of
spacetime in space and time, that is frame-dependent in special
relativity \cite{Minkowski1909}, is instead uniquely given at each
event. This foliation is an absolute one, an absolute physical
property intrinsic to the manifold. When gravitation is present,
by measuring the metric in principle one can decide whether a test
body is intrinsically at rest or not with respect to the manifold
under question.\par Solutions to the vacuum field equations of
general relativity that are static in the sense of Ehlers and
Kundt do exist. Solutions belonging to the class found by Weyl
\cite{Weyl1917} and by Levi-Civita \cite{Levi-Civita1919} have
been proved \cite{EK1964} to be static in that sense. It is
mandatory to interpret physically the static solutions of the
field equations of general relativity by paying due attention to
the peculiar uniqueness property exhibited by their Killing group.

\section{Choosing the manifold of Schwarzschild's
solution}\label{f} In general relativity, it is obvious that the
manifold to be associated to a given solution cannot be chosen by
deciding a priori the ranges of the coordinates in certain charts.
Since coordinates are mere labels, the choice of the manifold must
always rely on intrinsic physical arguments that need to be
developed a posteriori, once a solution to the field equations is
found. To this end the nontrivial Killing structure of the
solution needs to be investigated. Its outcome may be crucial for
the very choice of the manifold. As a corollary to the results of
the previous sections, the example of the intrinsically motivated
choice of the manifold that necessarily applies to the
Schwarzschild solution \cite{Schwarzschild1916},
\cite{Hilbert1917}, when the properties of its Killing group are
kept into account, is given here.\par In the symmetry-adapted
coordinates $x^1=r$, $x^2=\vartheta$, $x^3=\varphi$, $x^4=t$
chosen by Hilbert \cite{Hilbert1917}, the interval of
Schwarzschild's solution reads
\begin{equation}\label{6.1}
\dd s^2=\left(1-\frac{2m}{r}\right)\dd t^2
-\left(1-\frac{2m}{r}\right)^{-1}\dd r^2 -r^2(\dd {\vartheta}^2
+\sin^2{\vartheta}\dd \varphi^2),
\end{equation}
where $m>0$ agrees with the mass of a body in the Newtonian limit.
Due to the spherical Killing symmetry of the solution, we shall
set $-\pi/2 \le \vartheta \le \pi/2$ and $0 < \varphi \le 2\pi$.
When $r > 2m$, the solution is static in the sense of Ehlers and
Kundt. Like it occurs with the Weyl-Levi Civita solutions, at any
event a unique hypersurface-orthogonal wordline of absolute rest
is drawn, defined in the chosen coordinates through constant
values of $r$, $\vartheta$ and $\varphi$.  Let us consider a test
body on a certain worldline of the manifold, whose four-velocity
is $u^i\equiv\frac{\dd x^i}{\dd s}$; its acceleration four-vector,
i.e. the first curvature of its worldline \cite{Synge1960}, is
defined as
\begin{equation}\label{6.2}
a^i\equiv\frac{\DD u^i}{\dd s}\equiv\frac{\dd u^i}{\dd
s}+\Gamma^i_{kl}u^ku^l,
\end{equation}
where $\DD/\dd s$ indicates the absolute derivative. From it, one
builds the scalar quantity
\begin{equation}\label{6.3}
\alpha=(-a_ia^i)^{1/2}.
\end{equation}
When the test body lies on a worldline of absolute rest, the norm
of its four-acceleration, written in the chosen coordinates, reads
\begin{equation}\label{6.4}
\alpha=\left[\frac{m^2}{r^3(r-2m)}\right]^{1/2}.
\end{equation}
Through a given event, $\alpha$ is uniquely defined by the Killing
structure of the manifold \cite{EK1964}. Therefore it is clear
that this quantity, besides being an invariant, is intrinsic to
the manifold. In fact, due to the uniqueness of the
hypersurface-orthogonal time Killing vector, in the definition of
$\alpha$ no arbitrary choices based on elements foreign to the
structure of the manifold, like the arbitrary choice of a certain
worldline, have been invoked. One cannot accept that such an
invariant quantity intrinsic to the manifold may diverge somewhere
when approaching some event within the manifold. But $\alpha$
diverges when $r\rightarrow 2m$ from above. Hence the limiting
value of $r$ shall be $r>2m$, while in the limit $r\rightarrow
+\infty$ Newtonian physics is recovered. With these two choices
the range $2m< r<+\infty$ of the radial coordinate is therefore
fixed through intrinsic arguments. Then, of course
$-\infty<t<+\infty$. This choice of the manifold is in keeping
with the one done by Schwarzschild himself in his original work
\cite{Schwarzschild1916}.

\section{Further remarks on Schwarzschild's manifold}\label{g}
The existence of an intrinsic singularity when $r\rightarrow 2m$
due to the nontrivial Killing group structure of the solution is
decisive in the choice of the manifold done in the previous
Section. One should remember that, when the group-theoretical
argument recalled in the previous Section is overlooked and, like
it happened with the choice of the manifold done by Hilbert
\cite{Hilbert1917} in his reinterpretation of Schwarzschild's
original solution, the range of the radial coordinate $r$ is
assumed to be $0<r<+\infty$, a pathology soon appears. It
originates from the difference in the Killing groups prevailing
for $r>2m$ and for $r<2m$ respectively, and it can be avoided only
if the changes of topology of Hilbert's manifold produced by the
maximal extensions of Synge \cite{Synge1950}, Kruskal
\cite{Kruskal1960} and Szekeres \cite{Szekeres1960} are
introduced. This pathology is usually given scarce relevance in
the literature, although e.g. Rindler did not forget to mention it
in his book \cite{Rindler2001}. However, in order to appreciate
its full meaning, one has rather resorting to Synge, were a
detailed discussion of the issue of the time arrow in the
Schwarzschild solution can be found \cite{Synge1950}. In keeping
with Synge, a manifold meant to be a model of physical reality
must fulfill two postulates. One of them is the \textit{ postulate
of order}, according to which the parameter of proper time along a
timelike geodesic must always either decrease or increase; the
sense along which it is assumed to increase defines the sense of
the travel from past to future, namely the time arrow. Since the
geodesic equation (\ref{3.1}) is quadratic in the line element,
fixing the time arrow of the individual geodesic is a matter of
choice. The second postulate deals with our ideas of causation,
and establishes a relation between the time arrows of neighbouring
geodesics. Synge calls it the \textit{ non-circuital postulate}.
It asserts that \textit{ there cannot exist in space-time a closed
loop of time-like geodesics around which we may travel always
following the sense of the time-arrow}.\par Synge was the first to
show in detail \cite{Synge1950} that the time arrow can be drawn
in keeping with the aforementioned postulates in the maximally
extended manifold that he obtained from the Hilbert manifold with
his singular coordinate transformation; the same property
obviously holds in the Kruskal-Szekeres manifold too. Does it hold
also in the Hilbert manifold? A glance to Figure (1a) is
sufficient to answer the question in the negative. The arrow of
time can be drawn in keeping with Synge's two postulates of order
and of non-circuitality in the submanifold with $2m<r<+\infty$,
i.e. in Schwarzschild's original manifold, and separately in the
inner submanifold with $0<r<2m$. A consistent drawing of the arrow
of time, in keeping with both postulates, is however impossible in
Hilbert's manifold as a whole. This is an intrinsic flaw of the
latter manifold, originating from the abrupt change in the Killing
group structure that occurs at the crossing of the surface $r=2m$,
where the unique, hypersurface-orthogonal, timelike Killing vector
suddenly becomes spacelike. It has nothing to do either with the
fact that in Hilbert's chart the metric is not defined at $r=2m$,
or with the fact that in it the timelike geodesics appear to cross
the Schwarzschild surface at the coordinate time $t=\pm\infty$; it
is a flaw that cannot be remedied by \textit{ any} coordinate
transformation, however singular at $r=2m$, but one-to-one
elsewhere.\par The only known ways to overcome this flaw of
Hilbert's manifold are either by eliminating the inner region,
thereby reinstating the choice of the original manifold
\cite{Schwarzschild1916}, deliberately made by Schwarz\-schild as
a model for the gravitational field of a material particle, and
later confirmed by the study of the Killing group structure
\cite{AL2001}, or by completely renouncing the one-to-one
injunction on the coordinate transformations once set, on physical
grounds, by Einstein \cite{Einstein1916} and by Hilbert
\cite{Hilbert1915}.\par The second alternative is the one chosen
by Synge and his followers: in fact, not only Schwarzschild's
original manifold, but also Kruskal's manifold avoids the flaw of
the arrow of time present in Hilbert's manifold. Moreover, it
appears to preserve its inner region, for which $0<r<2m$. However,
it does so by a coordinate transformation that duplicates the
original manifold and alters its topology, in a way that is best
explained, rather than by looking at the equations for the
transformations, through a straighforward cut-and-paste procedure
applied to two Hilbert manifolds.\par
\begin{figure}[ht]

\includegraphics{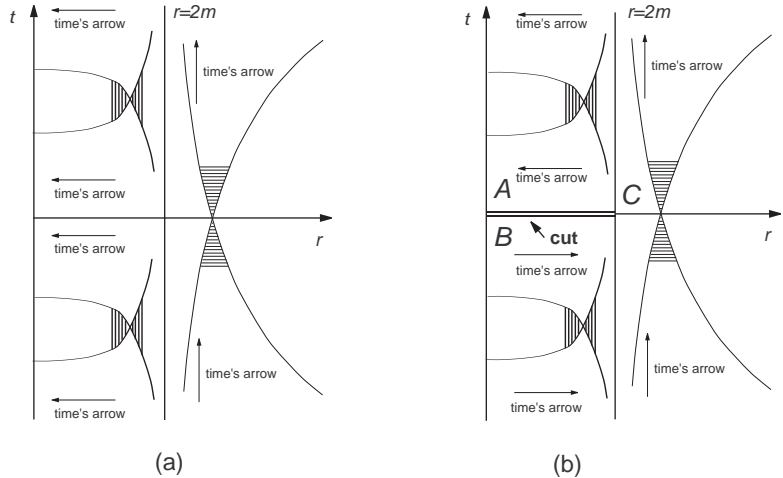}
\caption{\p (a): Drawing of Hilbert's manifold in the $r$, $t$
plane. Light cones are drawn both for $r<2m$ and for $r>2m$. Time
arrows are drawn in agreement with the non-circuital postulate.
Then the postulate of order happens to be violated. (b): Hilbert's
manifold is cut along $AC$. The topologically different manifold
obtained in this way allows for a drawing of the time arrow in
keeping with both Synge's postulates.}
\end{figure}

This procedure can be made in infinite ways, all entailing the
same change of topology. One of them is accounted for in the
sequence drawn in Figures (1b), (2a) and (2b) respectively.
\begin{figure}[ht]
\includegraphics{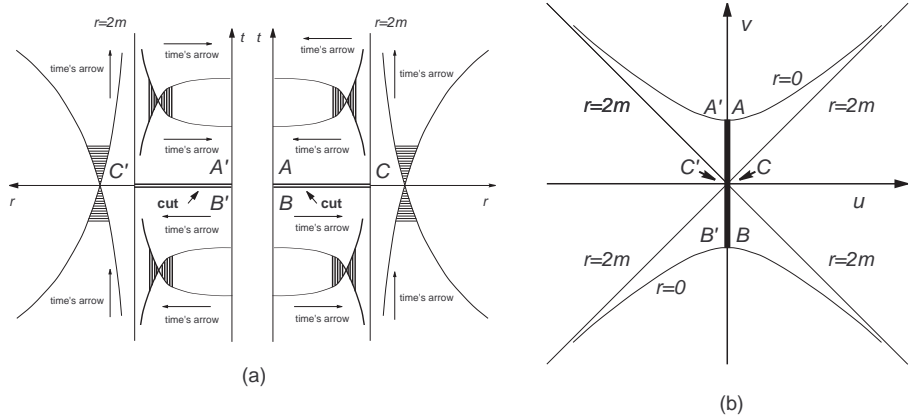}
\caption{\p (a): Two manifolds, equal to the Hilbert manifold cut
along  $ACB$ and  $A'C'B'$ in Figure (1b) are juxtaposed for
suturing; (b): sewing together the edges $ACB$ and $A'C'B'$
produces Kruskal's manifold, that fullfils both Synge's
postulates.}
\end{figure}
In Figure (1b) the inner region of the Hilbert manifold of Figure
(1a) is cut along the line $AC$ of the $r, t$ diagram. The
resulting manifold is topologically inequivalent to the Hilbert
manifold. The topological alteration already allows to draw the
arrows of time in keeping with both Synge's postulates but, due to
the existence of the border $ACB$, the new manifold is evidently
unphysical. However, if one takes two manifolds identical to the
one of Figure (1b), juxtaposes them as it is shown in Figure (2a),
and eventually sews together the borders $ACB$ and $A'C'B'$ like
in Figure (2b), one obtains a manifold equal to Kruskal's
manifold, and ascertains that the arrows of time inherited from
the two component manifolds with the cut still obey both Synge's
postulates.\par Only the topological alteration, drawn in the
Figures, that leads from the Hilbert manifold to the
 Kruskal
manifold  remedies the flaw of the time arrow due to the
coexistence, in the Hilbert manifold, of two submanifolds with a
different Killing group structure.

\section{Conclusion}\label{h}
The relation between relativity and invariance was clarified long
ago by the work of Felix Klein \cite{Klein1910},\cite{Klein1911}
and his result constitutes a paradigm that goes beyond the limits
of special relativity theory. Einstein's idea, that one should
introduce general transformations between curvilinear coordinates
was fundamental from a mathematical standpoint, since it allowed
one to avail of the powerful methods of the absolute differential
calculus of Ricci and Levi-Civita \cite{RLC1900}. However, the
physical idea by Einstein \cite{Einstein1916}, that certain
curvilinear coordinates should be availed of to account for
non-inertial reference frames, as required by the early form of
the equivalence principle, did not resist the criticism by
Kretschmann \cite{Kretschmann1917}. General curvilinear
coordinates are very useful mathematical tools, but they are
physically vacuous. The group of general covariance is physically
vacuous too. By following the ideas of Klein and Kretschmann, one
shall trace, in general relativity like in special relativity, the
physically meaningful group of invariance. The Killing group of
infinitesimal ``Mitschleppen'' of the metric tensor is such a
group. In a general solution of the field equations of 1915, the
Killing group is trivial, and general relativity then behaves like
an absolute theory. However, solutions with Killing groups that
are intermediate between the trivial one pertaining to an absolute
solution and the infinitesimal Poincar\'e group do exist. Their
scrutiny is fundamental for assessing the very structure of the
manifolds from a physical standpoint. When this scrutiny is
applied to the Schwarzschild solution, it turns out that only the
manifold originally chosen by Schwarzschild
\cite{Schwarzschild1916} survives. Other manifolds, like the ones
chosen by Hilbert \cite{Hilbert1917}, and later by Synge
\cite{Synge1950}, by Kruskal \cite{Kruskal1960} and by Szekeres
\cite{Szekeres1960} with their maximal extensions cannot survive
the scrutiny, because they contain a local, invariant, intrinsic
quantity that diverges in their interior.\par

\end{document}